\pdfoutput=1
\documentclass{article}
\usepackage{capt-of}
\usepackage{booktabs,caption,fixltx2e}
\usepackage[flushleft]{threeparttable}
\usepackage{graphicx}
\usepackage{makeidx} 
\usepackage{amsmath}
\usepackage{multirow}

\usepackage{amsfonts}%
\usepackage{amssymb}%
\usepackage[english,oldcommands,plain,lined]{algorithm2e}
\usepackage{graphics}
\usepackage{float}
\usepackage[labelfont=bf]{caption}
\usepackage[font={small}]{caption}
\usepackage{transparent}
\usepackage{color}
\usepackage{breqn}
\usepackage{epstopdf}
\usepackage{wrapfig}

\usepackage{color, colortbl}
\definecolor{LightCyan}{rgb}{0.88,1,1}
 \usepackage{cite}
\usepackage[font=small,labelfont=bf]{caption}
\usepackage{enumitem}
\usepackage{multirow}
\usepackage{cleveref}
\usepackage{makecell}
\usepackage{tikz}
 
\usepackage{pifont}
\newcommand{\cmark}{\ding{51}}%
\newcommand{\xmark}{\ding{55}}%
\usepackage{subfigure}

\begin{document}

\title{Multitask Learning of Temporal Connectionism in Convolutional Networks using a Joint Distribution Loss Function to Simultaneously Identify Tools and Phase in Surgical Videos} 

%
\author{Shanka Subhra Mondal \and Rachana Sathish \and Debdoot Sheet\\
Indian Institute of Technology, Kharagpur}
\date{}
%
%

\maketitle 
\vspace{-2mm}
\begin{abstract}

Surgical workflow analysis is of importance for understanding onset and persistence of surgical phases and individual tool usage across surgery and in each phase. It is beneficial for clinical quality control and to hospital administrators for understanding surgery planning. Video acquired during surgery typically can be leveraged for this task. Currently, a combination of convolutional neural network (CNN) and recurrent neural networks (RNN) are popularly used for video analysis in general, not only being restricted to surgical videos. In this paper, we propose a multi-task learning framework using CNN followed by a bi-directional long short term memory (Bi-LSTM) to learn to encapsulate both forward and backward temporal dependencies. Further, the joint distribution indicating set of tools associated with a phase is used as an additional loss during learning to correct for their co-occurrence in any predictions. Experimental evaluation is performed using the Cholec80 dataset. We report a mean average precision (mAP) score of $0.99$ and $0.86$ for tool and phase identification respectively which are higher compared to prior-art in the field.

\end{abstract}

\section{Introduction}
\label{sec:intro}
\begin{figure}
\includegraphics[width=\textwidth]{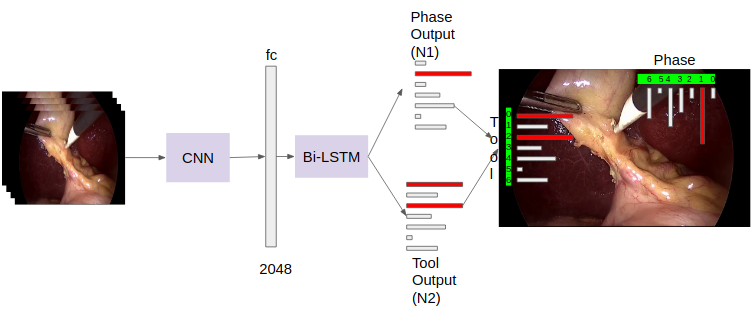}
\caption{The summary of the method  presented. The phase indices from 0-6 are in order \{Preparation, CalotTriangleDissection, ClippingCutting, GallbladderDissection, GallbladderPackaging, CleaningCoagulation, GallbladderRetraction\}. The tool indices from 0-6 are in order \{Grasper, Bipolar, Hook, Scissors, Clipper, Irrigator, SpecimenBag\}.
} \label{graphical_abstract}
\end{figure}

Surgical workflow analysis using videos acquired from an endoscope is of assistance to surgeons and hospital administrators to assess quality and progress of surgery and for medico-legal litigation. Being able to provide tool usage information during surgery along with its phase, report generation, determining the duration of surgery, time to completion of surgery are some of such useful information. This information summarizing also makes it easy to find out aberration in pattern of a particular tool usage during a surgery by comparing it with the reports of past procedures. This paper\footnote{Accepted paper at $5^{th}$ MedImage workshop of $11^{th}$ Indian Conference on Computer Vision, Graphics and Image Processing , Hyderabad, India, 2018} presents a multi-task deep learning framework which simultaneously infers both tool and phase information in video frames. The summary of the  method is presented in Fig.\ref{graphical_abstract}.

\textbf{Challenges:} In surgical videos the tools often appear occluded behind anatomical structures, which makes the task of tool detection difficult. Also the endoscope used to acquire video suffers from motion jitters leading to variation in scene background and the degree of illumination. Specular reflection is another related artifact. This makes the task of analysis typically challenging on account of the large scale vision appearance modeling to be performed. In a related note, this would also require training data to consist of large number of frames annotated with both tool and phase information, belonging to microcosms making up such wide-scale visual variations, which being a tedious job is also challenging to collect.

\textbf{Approach:} In the proposed framework a convolutional neural network (CNN)\cite{lecun1995convolutional} is trained simultaneously for both tool and phase detection where and it learns to extract high level visual features from the frames. Trained with an additional weighted joint distribution loss function which captures the joint probability of co-occurrence of a particular set of tools generally associated with a phase of the surgery. The visual features extracted from the trained CNN is used to train a bi-directional long short term memory network (LSTM)\cite{schuster1997bidirectional} to capture both the forward and backward temporal information across video frames. This temporal connectionism is important since in a surgery the phases are sequentially executed and there is an order in which a set of tools are used per phase. 

\textbf{Impact:} On account of introducing the weighted joint probabilistic loss function used in this multi-task training of CNN and bidirectional temporal learning with LSTM, the mean average precision for tools is higher than all the previous works related to this domain whereas in case of phase detection the mean average precision is comparable to the state of the art. The significant advancement over known prior-art in the field is the ability to use a single network to solve both phase and tool detection simultaneously, at the highest performance metric, achieved through its auto-correcting learning ability using joint distribution modeling. 

\textbf{Organization of the paper:} The earlier works on surgical tool and phase detection are briefly described in Sec. 2. The problem statement is presented
in Sec. 3. The methodology is explained in Sec. 4 . The experiments  are detailed with the results in Sec. 5. Sec. 6 presents the discussion. The conclusion is presented in Sec. 7 .

\section{Prior Work}
Various types of video analysis solutions have been proposed through the years. Use of 3D CNNs~\cite{ji20133d}, combining optical flow information along with 2D images ~\cite{simonyan2014two}, use of a RNN/LSTM along with a CNN to model long term dependencies~\cite{donahue2015long} are some of the widely known techniques. Many variants of these approaches have been used in both surgical phase and tool detection.

A CNN was trained to sort surgical video frames to learn temporal context between the frames and combined with a gated recurrent units (GRU) for surgical phase detection~\cite{bodenstedt2017unsupervised}. Later~\cite{twinanda2017endonet} proposed a multi-task CNN framework for both tool and phase detection, extracting the features from it and applying an hierarchical hidden Markov model (HHMM) for final phase detection. Another work~\cite{primus2018frame} used a CNN for phase classification in cataract surgery, and improved their accuracy by dataset purification and balancing. Later on~\cite{dergachyova2016automatic} constructed a surgical process modelling, and extracted various descriptors from images and then classified them using an Adaboost classifier. Further the temporal aspect was exploited using a hidden semi Markov model.
In~\cite{klank2008automatic} they proposed an evolutionary search in the space of global image features using a genetic programming based approach for phase detection in cholecystectomy videos. Another approach processes information about tool usage using non-visual electromagnetic tracking sensors and endoscopic camera for phase detection in laparoscopic surgeries using a left-right Hidden Markov Model (HMM)~\cite{padoy2008line}. Later~\cite{lalys2014surgical} proposed a framework to automatically detect surgical phases from microscope
videos. It first defined visual cues manually that can be helpful for discriminating the high-level tasks. The visual cues are automatically detected by image based
classifiers, and the obtained time series are then aligned with a reference surgery using dynamic time warping (DTW) algorithm for phase detection.
Successively~\cite{lea2016surgical} used a spatio-temporal CNN and also encoded tool and temporal information in it for extracting visual features from surgical frames and then built a classifier using DTW. In a prior work~\cite{mishra2017learning} tool presence in surgical video frames was detected by extracting visual features from a CNN and then feeding it to a LSTM for learning the temporal connectionism. Similar styled work~\cite{sznitman2014fast} proposed a tool detection system for minimally invasive surgery based on a multiclass ensemble classifier which was built using gradient boosted regression trees. Subsequently \cite{twinanda2017endonet} used only a CNN based approach for tool detection in each frame without considering the temporal information across video frames. Later \cite{speidel2009automatic} proposed an automatic method for detection of instruments from endoscopic images by segmenting the tip of the instrument and then recognizing based on three dimensional instrument models.
Earlier works in \cite{ryu2012endoscopic} used  image processing techniques like k-means clustering and Kalman filtering for localization and
tracking of tools in surgical videos. In \cite{sahu2016tool} combined  features extracted from pretrained and fine-tuned imagenet models to create contextual features for tool detection and later proposed a label set sampling to reduce the bias. Later \cite{al2017surgical} proposed to use optical flow information between surgical images to exploit spatial redundancies between consecutive images. Subsequently in \cite{al2018monitoring} proposed the CNN along with RNN framework followed by boosting of both of these networks and finally smoothing the predictions for surgical tool detection.
All of the methods described above for tool and phase detection use a CNN or a CNN + RNN framework or statistical methods, but none of those captures the joint probability distribution between the tools associated with a given phase while building a multitask learning framework. Also temporal information is captured in most of the works but they only consider the effect of past frames in determining the present tool or phase. It is equally important to look into the future as much as into the past for more accurate prediction in the current scenario and consider multitask framework in temporal domain.

\section{Problem Statement}

Given a video frame $\mathcal{F}^t$ it contains information about a particular phase of surgery and the multiple tools used which varies from a minimum of no-tool to a maximum of three tools. Given in a surgical video dataset, the ground truth for phase annotation in a frame is represented as a one-hot tensor $\mathbf{y}^t_1 \in{ \{0,1}\}$ of size $N_1\times 1$, where $N_1$ is the number of surgical phases. The surgical tools ground truth is represented as a multi-hot tensor $\mathbf{y}^t_2 \in{ \{0,1}\}$ of size $N_2\times 1$, where $N_2$ is the number of surgical tools. The prediction problem is modelled as $\{\hat{\mathbf{y}}^t_1, \hat{\mathbf{y}}^t_2\} \leftarrow \mathcal{H}\left(\mathcal{F}^t\right)$ where $\hat{\mathbf{y}}^t_1$ and $\hat{\mathbf{y}}^t_2$ are the phase and tool prediction tensors obtained from the trained multitask network $\mathcal{H}$ which processes $\mathcal{F}^t$. In case of tools none or more than one tool indices can be one in a given frame, so the detection of tools from a given video frame is a multilabel multiclass classification problem where we have to predict a subset of tools out of the total set of $N_2$ tools.  
\section{Exposition to the Solution}
\begin{figure}[h]
\includegraphics[width=\textwidth]{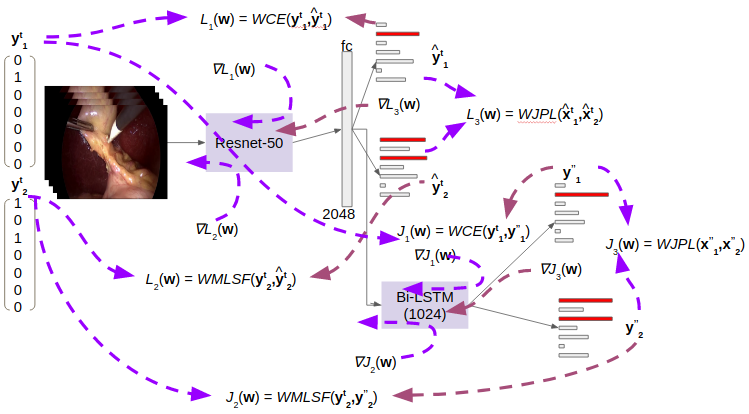}
\caption{The full training pipeline. The two one hot tensors(left) are the ground truths for phase and tool respectively. The arrows through the blocks of Resnet-50 and Bi-LSTM demonstrate the gradient flow. WCE, WMLSF, WJPL stands for weighted cross entropy loss, weighted multi label soft margin loss, weighted joint probabilistic loss respectively. Other arrows demonstrate which tensors contribute to the loss being computed. The bars denote the confidence levels of corresponding predictions with top predictions marked in red. }
\label{training_pipeline}
\end{figure}
We propose a multitask learning framework using CNN+LSTM to jointly solve for both tool and phase detection while learning with a weighted joint probability based loss function to model the dependence of tool and phase occurrence in a given frame. We \emph{first} train a CNN only with the multi-task setting. \emph{Second} we use the features from the penultimate fully-connected layer of the CNN trained earlier to construct a Bidirectional LSTM (Bi-LSTM) trained with a multi-task framework. The full training pipeline is shown in Fig.\ref{training_pipeline}. These stages are subsequently detailed.

\subsection{Multitask learning of a CNN for phase and tool detection}
\label{subsec:cnn}

Since the amount of tool annotated data is less, training a deep CNN architecture from scratch has been observed to lead to convergence challenges as well as slows down convergence. So to speed up the training process we have used a CNN trained prior on ImageNet for Large Scale Visual Recognition Challenge (ILSVRC)~\cite{deng2009imagenet}.
The ResNet-50 \cite{he2016deep} is used as a feature extractor and is finetuned on the task specific dataset after replacing the output layer. The input to the ResNet-50 is an image of size $224\times 224$ px and the features are obtained from the last but one fully connected layer of dimension $2,048$. The output layer in ResNet-50 is replaced to accommodate both tool and phase classifications with tensors matching properties of $\mathbf{y}^t_1$ and $\mathbf{y}^t_2$. 

Three different loss functions are used during training. During learning of \emph{phase detection}, the weighted cross entropy loss is used

\begin{equation}
\label{eq:toolloss}
L_1(\mathbf{y}^t_1,\hat{\mathbf{y}}^t_1)=w_1[n] \bigg(-y^t_1[n]+ \log\bigg(\sum_{j=0}^{N_1-1} \exp(\hat{y}^t_1[j])\bigg)\bigg)
\end{equation}

\noindent where $n=\arg\max \{\hat{y}^t_1[j]\forall j\in[0,N_1-1]\}$ with $\mathbf{y}^t_1=\{y^t_1[j]\forall j\in[0,N_1-1]\}$, and $w_1[j]$ is the weight associated with the $j^{th}$ phase out of the $N_1$ classes where the weight is obtained by median frequency balancing to compensate for high class imbalance in training data.

In case of \emph{tool detection} a weighted multi-label soft margin loss is used

\begin{eqnarray}
\label{eq:phaseloss}
L_2(\mathbf{y}^t_2,\hat{\mathbf{y}}^t_2)=\sum_{i=0}^{N_2-1}w_2[i]\hat{y}^t_2[i] \log\left(\frac{1}{(1+\exp(-y^t_2[i]))}\right) +\nonumber \\ w_2[i](1-\hat{y}^t_2[i]) \log\bigg(\frac{\exp(-y^t_2[i])}{1+\exp(-y^t_2[i])}\bigg)
\end{eqnarray}

\noindent where $\hat{y}^t_2[i]$ is the prediction of the $i^{th}$ tool in $\mathcal{F}^t$ and $y^t_2[i]$ is ground truth annotation for the tool presence with $\mathbf{y}^t_2=\{y^t_2[i]\forall i\in[0, N_2-1]\}$, $w_2[i]$ is the tool class weight obtained by median frequency balancing to compensate for high class imbalance in training data.

The third component of the loss takes in consideration the model of \emph{joint distribution of tool and phase occurrence} which is given as 

\begin{equation}
\label{eq:tool-phase-loss}
L_3(\hat{\mathbf{x}}^t_1,\hat{\mathbf{x}}^t_2)=\sum_{i=0}^{N_2-1}\sum_{j=0}^{N_1-1}\hat{x}^t_1[i]\hat{x}^t_2[j]IF(i,j)
\end{equation}

\noindent where $\hat{\mathbf{x}}^t_1=\sigma(\hat{\mathbf{y}}^t_1)$ and $\hat{\mathbf{x}}^t_2=\mathrm{SoftMax}(\hat{\mathbf{y}}^t_2)$ where $\sigma(\cdot)$ represents the sigmoid non-linearity, and $IF(i,j)$ denotes the inverse of the frequency of occurrence of tool $i\in[0,N_2-1]$ with a phase $j\in[0,N_1-1]$. Using the information present in the annotated training data we create a phase-tool co-occurrence matrix $\mathbf{C}=\{c_{i,j}\forall i\in[0,N_2-1], j\in[0,N_1-1]\}$ which represents the count of the number of frames over all videos when the $j^{th}$ tool was being used in the $i^{th}$ phase of surgery. Subsequently we form a normalized matrix $\hat{\mathbf{C}}=\{\hat{c}_{i,j}\forall i\in[0,N_2-1], j\in[0,N_1-1]\}$ with $\hat{c}_{i,j}=\frac{c_{i,j}}{\sum_{i=0}^{N_2-1}c_{i,j}}\forall j\in[0,N_1-1]$. This is used to create an IF function defined as $IF(i,j)=\frac{1}{\hat{c}_{i,j}+ \epsilon}$ where $\epsilon$ is the smallest value represented in the number system being used. This function is characterized such that if frequency of phase-tool co-occurrence turn out to be zero then a large value is represented in IF to induce a very high loss in that case.

\subsection{Multitask learning of a Bi-LSTM}
\label{subsec:bilstm}

The features extracted from the penultimate fully connected layer of the ResNet-50 trained earlier are used to train a multitask Bi-LSTM~\cite{ma2016end} in a similar learning framework using same cost functions as in (\ref{eq:toolloss}), (\ref{eq:phaseloss}) and (\ref{eq:tool-phase-loss}). Whitening transform~\cite{shental2002adjustment} is applied to all features across the training data being fed to the Bi-LSTM. Due to its bidirectional nature it maintains two hidden layers, where ones propagates from left to right in the time unrolled sequence, and the other from right to left. The final classification result, is generated through
combining the score results produced by both the LSTM hidden layers. The input to the bidirectional LSTM is  sequence of visual features from the frames extracted from the entire video. A single layered Bi-LSTM with $1,024$ hidden neurons was used. Finally median filtering is applied to the phase predictions to remove any abrupt changes. 

\section{Experiments and Results}

\subsection{Dataset Description}

\begin{table}[t]
\centering
\caption{Mean $\pm$ Standard Deviation of duration for seven different phases in Cholec80 dataset}
\label{table:phase_dur}
\begin{tabular}{|c|c|c| }
 \hline
 \multicolumn{1}{|c|}{Phase Id}&
 
\multicolumn{1}{c|}{Phase Name} & \multicolumn{1}{c|}{Duration (secs)} \\
 \hline
 
  P1  & Preparation  & $ 125\pm95$\\
 \hline
  P2  & Calot triangle dissection  & $ 954\pm538$\\
 \hline
  P3  & Clipping and cutting  & $ 168\pm152$\\
 \hline
  P4  & Gallbladder dissection  & $ 857\pm551$\\
 \hline
  P5  & Gallbladder packaging  & $ 98\pm53$\\
 \hline
  P6  & Cleaning and coagulation  & $ 178\pm166$\\
 \hline
  P7  & Gallbladder retraction  & $ 83\pm56$\\
 \hline

 \end{tabular}
 \end{table}

\begin{figure}[h]
\includegraphics[width=\textwidth]{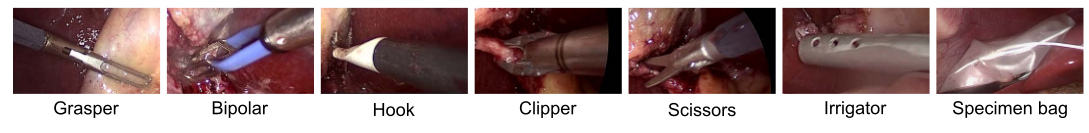}
\caption{List of seven different surgical tools present in the Cholec80 dataset.}
\label{tools}
\end{figure}

\begin{figure}
    \centering
    \subfigure[Phase occurrence]{\includegraphics[width=0.5\textwidth]{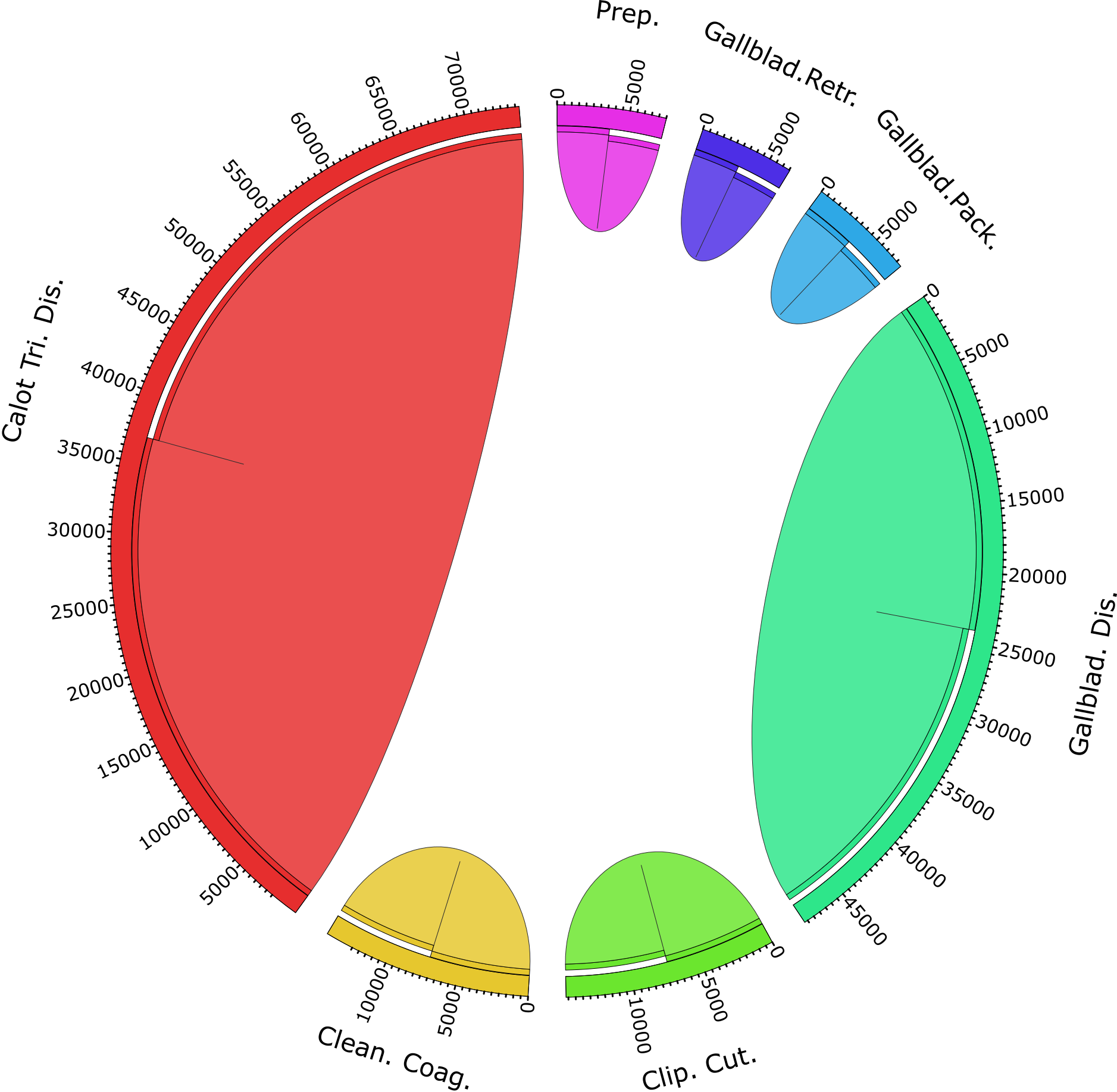}\label{phase}}%
    \subfigure[Tool co-occurrence]{\includegraphics[width=0.5\textwidth]{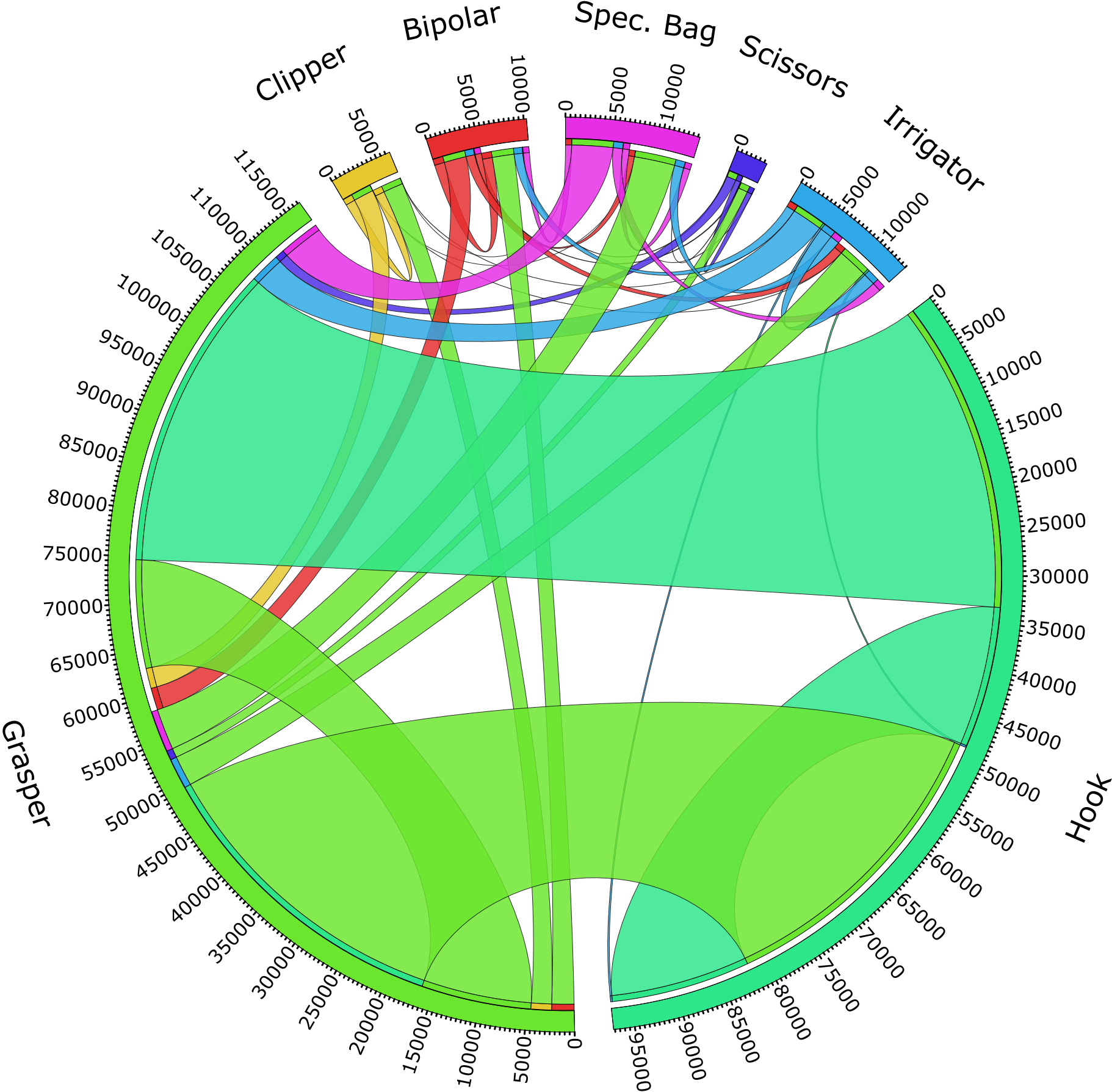}\label{tool}}%
    \caption{Tool and Phase Distribution in Cholec80 training dataset.}%
    \label{tool_phase}
\end{figure}

\begin{figure}[H]
\includegraphics[width=\textwidth]{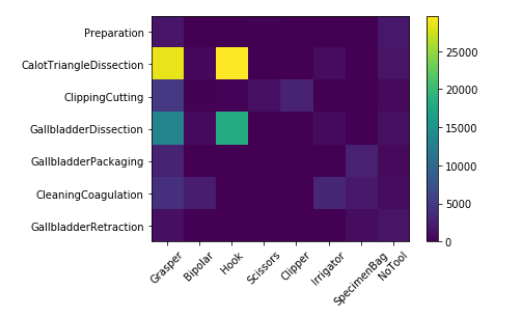}
\caption{Co-occurrence matrix of tools and phase in Cholec80 training dataset.}
\label{tool_phase_heatmap}
\end{figure}

The proposed method is evaluated on Cholec80\footnote{\url{http://camma.u-strasbg.fr/datasets}} dataset which contains 80 videos of cholecystectomy surgeries performed by 13 surgeons at the University Hospital of Strasbourg. The phase annotation is provided for all the frames at 25 frames per second (fps) whereas tools are annotated on one per 25 frames leading to 1 fps annotation rate on a 25 fps video. These annotations are rate matched to 1 fps. The dataset is split into two equal parts, the first 40 videos are used for training the multitask CNN and Bi-LSTM and the last 40 videos are used for validation or testing. The visual appearance and list of 7 surgical tools in Cholec 80 dataset is given in Fig.\ref{tools}. The details about the seven different surgical phases and the mean$\pm$std of their duration in given in Table. ~\ref{table:phase_dur}. Also the dataset is imbalanced with respect to both surgical phases and tools as evident in Fig.~\ref{phase} and Fig.~\ref{tool} respectively.  Accordingly $N_1=7$ corresponding to 7 phases of surgery and $N_2=8$ corresponding to 7 tools and the no-tool case. The phase-tool co-occurrence matrix $\mathbf{C}$ can be visualized in Fig.~\ref{tool_phase_heatmap}. 

\subsection{Training}

The \emph{multitask CNN} (Sec.~\ref{subsec:cnn}) is trained with a learning rate of $1 \times 10^{-4}$ with a learning rate scheduler which reduces the learning rate by $0.9$ when the validation loss did not decrease for more than $5$ consecutive epochs of training, batch size of $100$ frames used, weight decay of $5\times 10^{-4}$, momentum of $0.9$. The network is optimized using stochastic gradient descent algorithm (SGD). 

The \emph{multitask Bi-LSTM} (Sec.~\ref{subsec:bilstm}) is trained with a learning rate of $1 \times 10^{-2}$ with a learning rate scheduler which reduces the learning rate by $0.5$ when the validation loss does not decrease for more than $5$ epochs consecutive during training, batch size of $1$ video is used, and remaining parameters as same.

\subsection{Baselines}

For comparison of the performance of the proposed method we have considered seven baselines. BL1 is the modified multi-label multi-class Resnet-50 which predicts tools present on an individual frame without using any temporal information in videos. BL2 is BL1 along with Bi-LSTM. BL3 is modified multi-class ResNet-50 used only for phase prediction using individual frame. BL4 is BL3 + Bi-LSTM. BL5 is modified ResNet-50 and it jointly predicts both tool and phase on individual frames only and trained using the 3 loss functions. BL6 is Endonet~\cite{twinanda2017endonet} which predicts both tool and phase. BL7 is boosted CNN + RNN~\cite{al2018monitoring} which predicts only tool. The \emph{proposed method} is essentially BL5 + Bi-LSTM.

\subsection{Implementation}

The experiments were implemented using PyTorch 0.4\footnote{https://pytorch.org} and accelerated with Nvidia CUDA 9.0\footnote{https://developer.nvidia.com/cuda-90-download-archive} and cuDNN 7.3\footnote{https://developer.nvidia.com/cudnn} on Ubuntu 16.04 LTS Server OS. The server consisted of 2x Intel Xeon E5-2699 v3 CPU, 2x32 GB DDR4 ECC Regd. RAM, 4TB HDD, 1x Nvidia Quadro P6000 GPU with 24 GB DDR5 RAM. The CNN models (BL1, BL3, BL5) were trained for $200$ epochs while the Bi-LSTM for the adjunct models (BL2, BL4, Proposed method) for $1,000$ epochs.

\subsection{Results}

\begin{figure}[!htbp]
\centering
\includegraphics[width=0.7\textwidth]{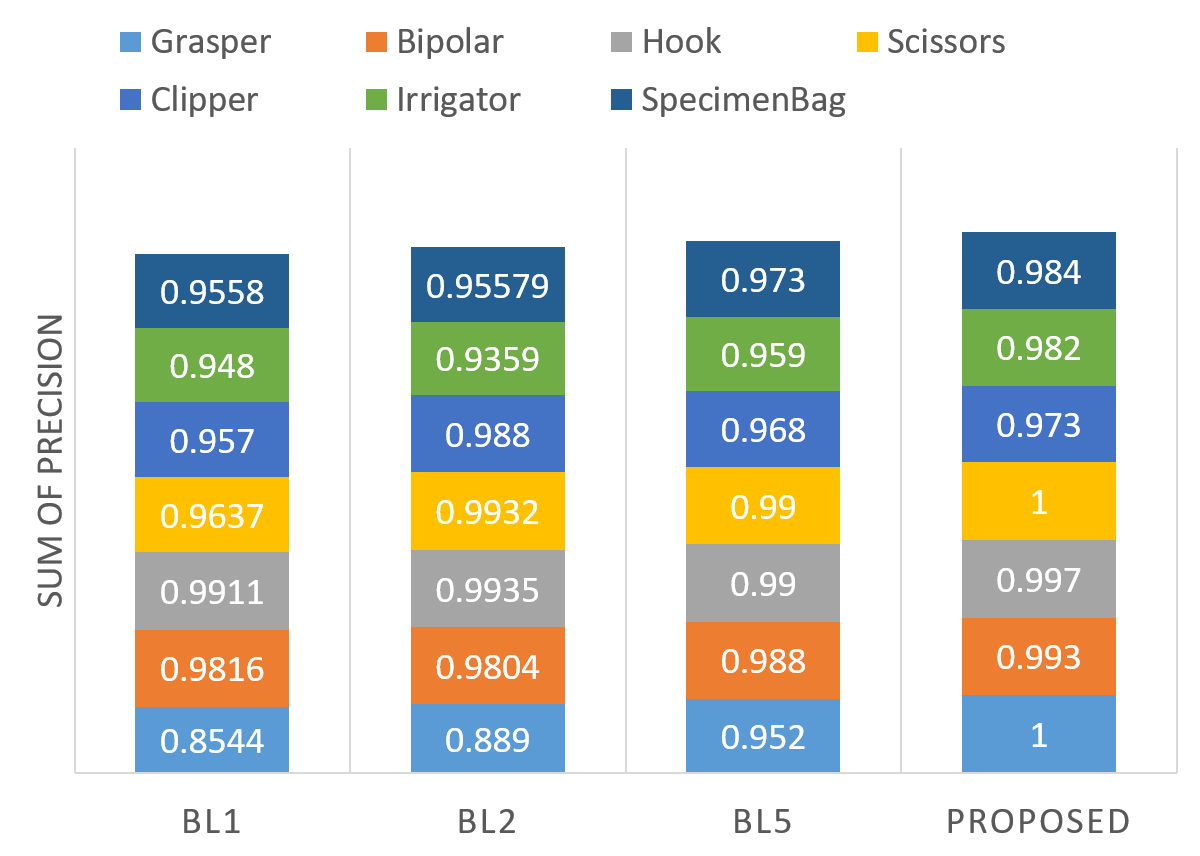}
\caption{Performance of BL1, BL2, BL5 and proposed method for tool precision}
\label{tool_precision}
\end{figure}

\begin{figure}[h]
\centering
\includegraphics[width=0.7\textwidth]{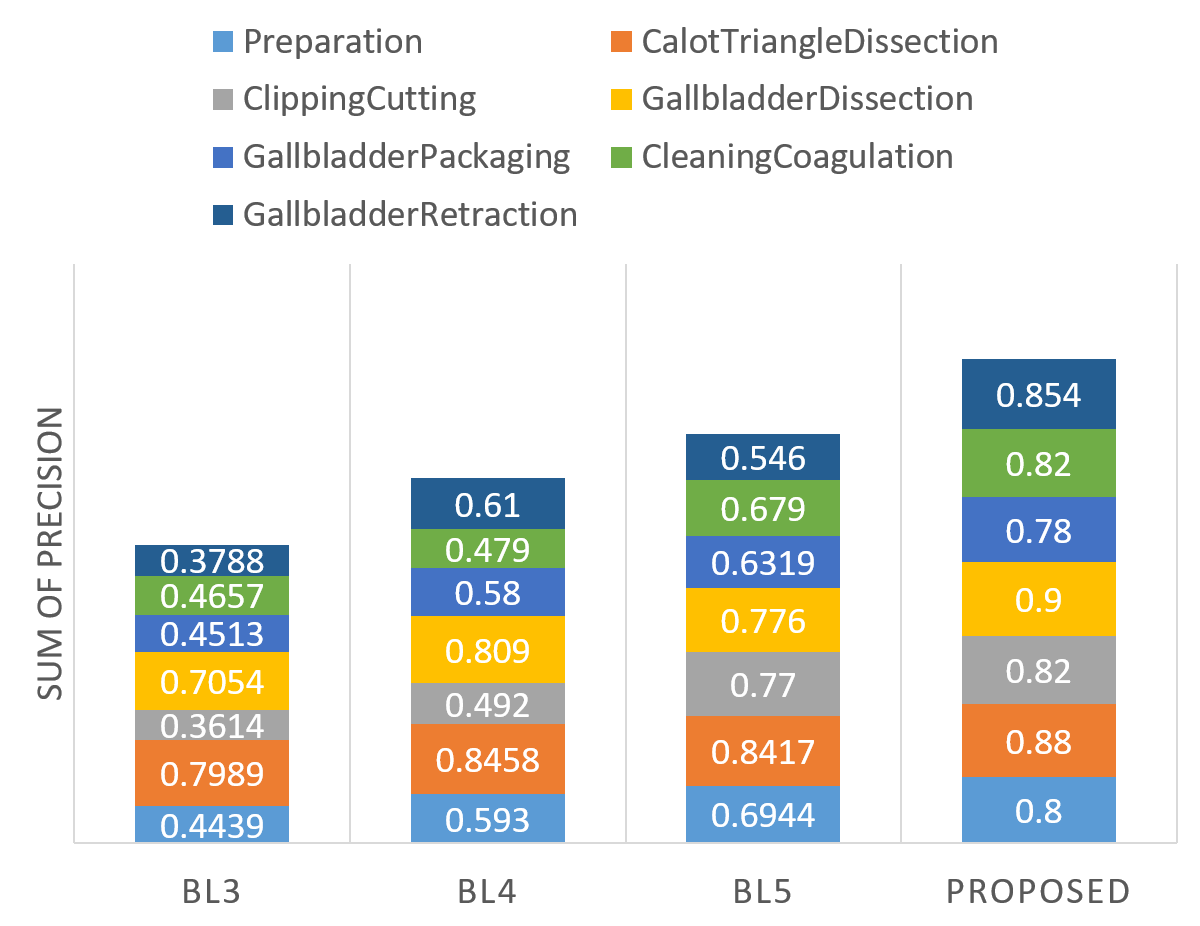}
\caption{Performance of BL3, BL4, BL5 and proposed method for phase wise precision}
\label{phase_precision}
\end{figure}

\begin{figure}[h]
\centering
\includegraphics[width=0.7\textwidth]{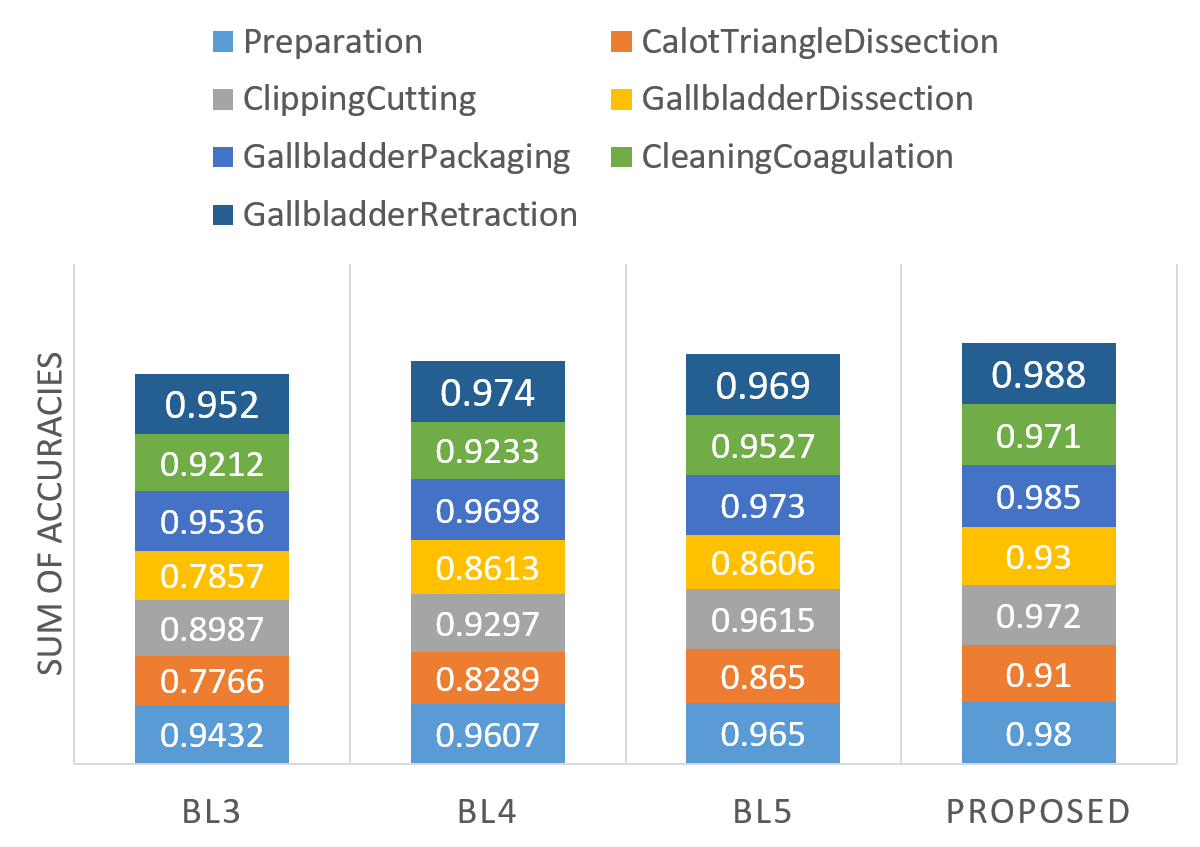}
\caption{Performance of BL3, BL4, BL5 and proposed method for phase wise accuracy}
\label{phase_acc}
\end{figure}

The comparison between the baselines and the proposed method for the three metrics namely average precision, average recall, average accuracy is shown in Table. \ref{table:results}. The performance of the baselines (BL1, BL2, BL5) and the proposed method for tool- wise precision is shown in Fig.\ref{tool_precision}. The performance of the baselines (BL3, BL4, BL5) and the proposed method for phase- wise precision and accuracy are shown in Fig.\ref{phase_precision} and Fig.\ref{phase_acc} respectively. All results are provided for the validation set (last 40 videos of Cholec80 dataset).

    

\begin{table}
\centering
\caption{Performance comparison of proposed method with baselines. Best performance metric indicated in bold face.}
\label{table:results}
\begin{tabular}{|c|c|c|c|c|c|c|c|c| }
 \hline
 \multicolumn{1}{|c|}{}&
 
\multicolumn{4}{c|}{Tool Detection} & \multicolumn{4}{c|}{Phase Detection} \\
 \hline
 
  Baseline & Tool  & \thead{Average \\Precision}  & \thead{Average \\Recall} &  \thead{Average \\Accuracy} & Phase  & \thead{Average \\Precision}  & \thead{Average \\Recall} &  \thead{Average \\Accuracy}\\
 \hline
 BL1   & \cmark&0.955    & 0.928 &   0.958 &\xmark& -&-&-\\
 \hline
 BL2   & \cmark&0.963    & \textbf{0.936} &   \textbf{0.964} &\xmark& -&-&-\\

 \hline
 BL3   & \xmark&-    & - &  - &\cmark& 0.515&0.63&0.88\\
 \hline
 BL4   & \xmark&-    & - &  - &\cmark& 0.63&0.717&0.92\\
 \hline
 BL5   & \cmark&0.974    & 0.88 &   0.938 &\cmark& 0.705&0.6944&0.935\\
 \hline
 BL6   & \cmark&0.81    & - &   - &\cmark& 0.848&\textbf{0.883}&0.92\\
 \hline
 BL7   & \cmark&0.9789    & - &  - &\xmark& -&-&-\\
 \hline
 \thead{Proposed \\Method}   & \cmark&\textbf{0.99}    & 0.912 &   0.9353 &\cmark& \textbf{0.857}&0.835&\textbf{0.966}\\

 \hline

\end{tabular}
\end{table}

\section{Discussion}

In this paper we have proposed a new loss function for multitask learning using a weighted joint probabilistic loss function to model the dependency of a set of tools to a phase in laparoscopic surgeries. Subsequently we use CNN and Bi-LSTM framework which jointly predicts tool and phase. We show through experiments that the mean average precision (mAP) obtained for tool detection outperforms all other previous architectures. In case of phase detection it yields better results with respect to mAP and also yields a higher accuracy. This indicates that the visual features learned by the CNN provides valuable information through rich features to the Bi-LSTM. Also the interdependence between tool and phase provided to the network through the weighted joint probabilistic loss function, which ultimately affects gradients and update of parameters helps in better convergence. Another important aspect of our framework is the use of Bi-LSTM, which has an inherent capability to capture long term dependencies both along past and future, expected to be required for better prediction in temporal domain. In Bi-LSTM full video batch stacking and whitening transform of CNN features prior to learning yield significantly better performance and faster convergence. Also the median filtering applied to the phase predictions obtained from Bi-LSTM resulted in slight improvement in mAP and accuracy due to  removal of abrupt changes.

The results are provided for Cholec80 dataset which contains 80 videos of cholecystectomy surgeries. Some of the previous works have used less than 20 videos of surgeries for surgical work-flow analysis which had limited their performance on account of its inability to learn the richness of visual appearances associated with tools and phases. Without using any data augmentation techniques to compensate for the tool and phase imbalance as seen from Fig.\ref{phase} and Fig.\ref{tool} the model gave significantly better results, which suggests that it is robust to  data imbalance. The dataset also contains lot of variability with respect to phase duration as seen from Table. \ref{table:phase_dur} which does not affect  the phase detection results to any significant extent thereby demonstrating the network's capability to tackle such challenges. 

Although the model can overcome the challenges described above there are some limitations. Firstly, Cholec80 dataset is limited to surgeons from one institution and can easily lead to over-fitting and hence a dataset containing surgeries from multiple surgeons from different institutions should be used for training which can yield more generalized results. Secondly, no image processing techniques were applied to the raw frames extracted from videos to remove redundant information which can help the CNN to learn better features. Thirdly the framework requires training of the CNN first followed by a Bi-LSTM, while making it as an end to end system would require training only once which would be less computationally expensive and is desired.

\section{Conclusion}

A multitask deep learning framework comprised of ResNet-50 and Bi-LSTM with a weighted joint distribution loss function has been proposed. It gives better mAP with respect to tool detection and comparable results for phase detection. The applicability of the proposed method is not necessarily limited only to tool and phase detection but other areas such as tool localization, estimating completion time of surgery, recognition of anatomy should be explored. Also the tools in many images can have various orientations with respect to the camera depending on the surgery, so the use of vector convolutions~\cite{marcos2017rotation} can make the system rotation invariant which can be seen as a future work to improve tool and phase prediction with ability to learn with limited annotated data corpus.



\end{document}